\documentclass[usegraphicx,usenatbib]{mn2e}

\begin{document}
\title[Primordial Non-Gaussianity from WMAP]{Limits on Primordial
Non-Gaussianity from Minkowski Functionals of the WMAP Temperature
Anisotropies}

\author[Hikage et
  al.]{C.~Hikage$^{1,2}$\thanks{chiaki.hikage@astro.cf.ac.uk},
  T.~Matsubara$^3$, P.~Coles$^1$, M.~Liguori$^4$, F.~K.~Hansen$^5$,
  S.~Matarrese$^{6,7}$ \\
%\newauthor
$^1$
  School of Physics and Astronomy, Cardiff University,
  Queens Buildings, 5, The Parade, Cardiff CF24 3AA \\
$^2$
  School of Physics and Astronomy, University of Nottingham,
  University Park, Nottingham, NG7 2RD \\
$^3$
  Department of Physics and Astrophysics,
  Nagoya University, Chikusa, Nagoya 464-8602, Japan \\
$^4$
  Department of Applied Mathematics and Theoretical Physics,
  Centre for Mathematical Sciences, University of Cambridge, \\
  Wilberfoce Road, Cambridge, CB3 0WA \\
$^5$
  Institute of Theoretical Astrophysics, University of Oslo,
  P.O. Box 1029 Blindern, 0315 Oslo, Norway \\
$^6$
  Dipartimento di Fisica ``G. Galilei'' universit$\acute{\rm a}$
  di Padova, INFN Sezione di Padova,
  via Marzolo 8, 1-35131 Padova, Italy \\
$^7$
 INFN, Sezione di Padova, via Marzolo 8, I-35131, Padova, Italy
}

%\date{\today}
\date{Submitted 2008 February 25; Accpeted 2008 July 2}
%\pagerange{\pageref{firstpage}--\pageref{lastpage}} \pubyear{2002}
\maketitle
%%%%%%%%%%%%%%%%%%%%%%%%%%%%%%%%%%%%%%%%%%%%%%%%%%%%%%%
\begin{abstract}
We present an analysis of the Minkowski Functionals (MFs) describing
the WMAP three-year temperature maps to place limits on possible
levels of primordial non-Gaussianity. In particular, we apply
perturbative formulae for the MFs to give constraints on the usual
non-linear coupling constant $f_{\rm NL}$. The theoretical
predictions are found to agree with the MFs of simulated CMB maps
including the full effects of radiative transfer. The agreement is
also very good even when the simulation maps include various
observational artifacts, including the pixel window function, beam
smearing, inhomogeneous noise and the survey mask. We find
accordingly that these analytical formulae can be applied directly
to observational measurements of $f_{\rm NL}$ without relying on
non-Gaussian simulations. Considering the bin-to-bin covariance of
the MFs in WMAP in a chi-square analysis, we find that the
primordial non-Gaussianity parameter is constrained to lie in the
range $-70<f_{\rm NL}<91$ ($95\%$ C.L.) using the Q+V+W
co-added maps.
\end{abstract}

%%%%%%%%%%%%%%%%%%%%%%%%%%%%%%%%%%%%%%%%%%%%%%%%%%%%%%%
\begin{keywords}
Cosmology: early Universe -- cosmic microwave background
-- methods: statistical -- analytical
\end{keywords}
%%%%%%%%%%%%%%%%%%%%%%%%%%%%%%%%%%%%%%%%%%%%%%%%%%%%%%%

\section{Introduction}
The existence of non-Gaussianity in primordial density fields has
the potential to provide a unique observational probe that will
enable discrimination among wide variety of inflationary models of
the early Universe. Versions of the inflation scenario based on the
idea of a single slow-rolling scalar field predict levels of
non-Gaussianity too small to be observed. On the other hand,
multi-field inflation models and models with a non-standard kinetic
term for the inflaton may yield larger non-Gaussian effects which
could in principle be detected in current or next-generation
observations
\citep[e.g.][]{BMR2002,BU2002,lyth2003,dvali2004,ACMZ2004,AST2004,BKMR2004,Chen2007,BB2007,KMVW2007}.

In this paper we focus on the local parametrisation of primordial
non-Gaussianity by including quadratic corrections to the curvature
perturbation during the matter era \citep[e.g. ][]{KS2001}:
%%%%%%%%%%%%%%%%%%%%%%%%%%%%%%%%%%%%%%%%%%%%%%%%%%%%%%%
\begin{equation}
\label{eq:ngpotential2}
\Phi=\phi+f_{\rm NL}(\phi^2-\langle\phi^2\rangle),
\end{equation}
%%%%%%%%%%%%%%%%%%%%%%%%%%%%%%%%%%%%%%%%%%%%%%%%%%%%%%%
where $\phi$ represents an auxiliary random-Gaussian field and $f_{\rm
NL}$ characterizes the amplitude of the non-linear contribution to the
overall perturbation. This local form is motivated by the simple
slow-rolling single scalar inflation scenario and other models,
including curvaton models; for an alternative parameterization of
$f_{\rm NL}$, see \citep{Creminelli2007a}.  Current observations are
not sufficiently sensitive to detect the wavelength dependence of
$f_{\rm NL}$ so a constant $f_{\rm NL}$ provides a reazonable
parameterization of the level of non-Gaussianity.

Analysis of the angular bispectrum for the WMAP 3-year data provides a
constraint on $f_{\rm NL}$ to lie between $-54$ and $114$ at the
$95\%$ C.L. \citep{Komatsu2003,Spergel2007}.
\citet{Creminelli2007a} obtains more stringent constraint $-36<f_{\rm
NL}<100$. On the other hand, \citet{YW2008} recently reported a
detection of primordial non-Gaussianity at greater than 99.5\%
significance. Further detailed analyses of non-Gaussianity is clearly
necessary in order to reconcile and understand the various constraints
and claimed detections.

Different approaches to the study of non-Gaussianity exploit
different statistical properties and will be sensitive to different
aspects of the behaviour of the pattern being tested. In general,
there is no unique statistic to describe the non-Gaussian nature of
a sample in a complete manner. A given method may have strong
discriminatory power for one particular form of non-Gaussianity, but
this is not necessarily the case for all possible alternative
distributions. Testing non-Gaussianity therefore requires a battery
of complementary techniques rather than a single approach. It is
particularly important to use different statistical approaches in
the context of primordial non-Gaussianity, because the physical
mechanism responsible remains unknown. Furthermore, in the real
world, issues including survey masks and inhomogeneous noise have to
be taken into consideration. Different statistics may be sensitive
to different systematics and foreground artefacts, so that
complementary analysis using different statistics are essential for
a robust detection. Analyses using different statistical methods are
useful to validate or refute basic theoretical models and constrain
model parameters more accurately. The most commonly used statistic
for non-Gaussian analysis is the bispectrum (or even trispectrum)
which focuses on information contained within three-point (or
four-point) correlations. Other approaches represented by Minkowski
Functionals (MFs) and genus statistics (one of MFs) utilize
information concerning the integrated morphology and topology of the
density structure, and are dependent on all order of correlation
functions. Their robustness and generality therefore makes them
ideal complements to standard correlation analyses.

In this analysis, we focus on the {\em local} model of primordial
non-Gaussianity characterized by $f_{\rm NL}$ in Equation
(\ref{eq:ngpotential2}). \citet{Creminelli2007b} show that the
bispectrum is the optimal statistics in the estimation of $f_{\rm
NL}$ and then other statistics (e.g., trispectrum) are useless even
if there are different foreground contaminations. This is, however,
only the case when the local model exactly describes the real
universe. Other forms of non-Gaussianity, different from the one
purely characterized by $f_{\rm NL}$, which may exist in a real
observation, could make the fit of the theoretical estimation as a
function of $f_{\rm NL}$ to the observation worse and also influence
the estimation of $f_{\rm NL}$ among different statistics. Different
statistical approaches are, therefore, still useful to test the
assumed model of primordial non-Gaussianity and also possible
observational systematics by checking if they have a reasonable
goodness of the fit to observations and thus give a consistent limit
on $f_{\rm NL}$ compared to that that from the bispectrum.

In this paper, we present a measurement of primordial
non-Gaussianity from the MFs of the WMAP three-year temperature
maps. We apply perturbative formulae recently derived by
\cite{HKM2006} to do the comparison with observations;  previous
analyses rely on  non-Gaussian simulations
\citep{Komatsu2003,Spergel2007}.  The agreement of the theoretical
predictions with non-Gaussian simulations has already been
established in Sachs-Wolfe limit \citep{HKM2006}. In this paper, we
apply non-Gaussian simulations based on full radiative transfer
computations and then demonstrate that the analytical predictions
accurately reproduce the simulation results. \citet{Gott2007}
already derived an analytical formula for the genus statistic in the
Sachs-Wolfe approximation to compare with WMAP data. Our analysis
takes more detailed physics into account and is consequently more
accurately applicable to a wider range of scales.

Observational effects (including antenna beam pattern, inhomogeneous
noise and the survey mask) could be other sources of confusion. From a
comparison with simulations including these observational issues, we
find that the observational systematics are negligible to estimate the
primordial non-Gaussianity from WMAP data directly using our method.

The organization of this paper is as follows. In \S\ref{sec:data}
the WMAP three-year data studied here are briefly introduced. In
\S\ref{sec:comp_sim} we test whether the perturbative formulae well
describe the MFs for the non-Gaussian simulation maps even including
the various observational effects mentioned above. In
\S\ref{sec:wmaplimit}, we show the MFs for WMAP three-year
temperature maps compared with theoretical formulae and give
constraints on $f_{\rm NL}$. \S~\ref{sec:summary} is devoted to a
summary and the conclusions.

\section{WMAP Three-Year Data}
\label{sec:data}

The CMB temperature maps derived from the WMAP observation are
pixelized in HEALPix format with the total number of pixels $n_{\rm
pix}=12N_{\rm side}^2$ \citep{Gorski2005}. In our analysis, we use
the maps for Q, V and W frequency bands with $N_{\rm side}=512$. The
linearly co-added maps are constructed using an inverse weight of
the pixel-noise variance $\sigma_0^2/\bar{N}_{\rm obs}$, where
$\sigma_0$ denotes the pixel noise for each differential assembly
(DA) given in \cite{WMAPfg} and $\bar{N}_{\rm obs}$ represents the
full-sky average of the effective number of observations per each
pixel.  We adopt two maps with different combinations of frequency
bands: V and W (written as ``V+W'') and Q, V and W (written as
``Q+V+W'').  The co-added maps are masked with the {\it Kp0} galaxy
mask including point-source mask provided by \cite{WMAPfg}, which
leaves $76.8\%$ of the sky available for the data analysis.

In comparison with WMAP observations to give constraint on $f_{\rm NL}$
in \S\ref{sec:wmaplimit}, a $\Lambda$CDM cosmology is assumed with the
cosmological parameters at the maximum likelihood peak from the WMAP
three-year data only fit \citep{Spergel2007}: $\Omega_b=0.04309$,
$\Omega_{\rm cdm}=0.211$, $\Omega_\Lambda=0.74591$, $H_0=71.227~{\rm
km~s^{-1}~Mpc^{-1}}$, $\tau=0.08982$, and $n_s=0.95537$. The
amplitude of the primordial fluctuations has been normalized by the
first acoustic peak of the temperature power spectrum,
$l(l+1)C_l/(2\pi)=5617.05({\mu}{\rm K})^2$ at $l=220$
\citep{WMAP3peak}.

\section{Perturbative Formulae versus
Non-Gaussian Simulations}
\label{sec:comp_sim}

\subsection{Perturbative Formulae of MFs for CMB with Primordial Non-Gaussianity}

The topology of random fluctuation fields is generally studied using
their excursion sets, i.e. regions where the field exceeds some
threshold level. In a two-dimensional random field such as a CMB
temperature map, three MFs are defined: the fraction of area $V_0$
exceeding the threshold, the total circumference $V_1$ of all the
entire excursion set, and the corresponding Euler Characteristic
$V_2$ \citep{Coles88}. We measure MFs for CMB temperature maps as a
function of the threshold density $\nu$, defined as the temperature
fluctuation $\Delta T/T$ normalized by its standard deviation
$\sigma_0\equiv\langle(\Delta T/T)^2\rangle^{1/2}$.  Based on the
general formalism of perturbation theory for MFs
\citep{Matsubara2003}, \citet{HKM2006} derived perturbative formulae
of the MFs as a function of the non-linear coupling parameter
$f_{\rm NL}$ (eq.[\ref{eq:ngpotential2}]).

The MFs are separately written with the amplitude and the function
of $\nu$ as follows.
%%%%%%%%%%%%%%%%%%%%%%%%%%%%%%%%%%%%%%%%%%%%%%%%%%%%%%%
\begin{equation}
\label{eq:mf}
V_k(\nu)=A_k v_k(\nu).
\end{equation}
%%%%%%%%%%%%%%%%%%%%%%%%%%%%%%%%%%%%%%%%%%%%%%%%%%%%%%%
The amplitude $A_k$, which is determined only by
the angular power spectrum $C_l$, is given by
%%%%%%%%%%%%%%%%%%%%%%%%%%%%%%%%%%%%%%%%%%%%%%%%%%%%%%%
\begin{equation}
\label{eq:mfamp}
A_k=\frac1{(2\pi)^{(k+1)/2}}\frac{\omega_2}{\omega_{2-k}\omega_k}
\left(\frac{\sigma_1}{\sqrt{2}\sigma_0}\right)^k,
\end{equation}
%%%%%%%%%%%%%%%%%%%%%%%%%%%%%%%%%%%%%%%%%%%%%%%%%%%%%%%
\begin{equation}
\label{eq:var}
\sigma_j^2\equiv \frac1{4\pi}\sum_l(2l+1)\left[l(l+1)\right]^j C_l W^2_l,
\end{equation}
%%%%%%%%%%%%%%%%%%%%%%%%%%%%%%%%%%%%%%%%%%%%%%%%%%%%%%%
where $\omega_k\equiv \pi^{k/2}/{\Gamma(k/2+1)}$ gives $\omega_0=1$,
$\omega_1=2$, $\omega_2=\pi$ and $W_l$ represents the smoothing
kernel determined by the pixel and beam window functions and any
additional smoothing (e.g. a Gaussian kernel).  In weakly
non-Gaussian fields, the function $v_k(\nu)$ can be divided into the
Gaussian term $v_k^{(G)}$ and the non-Gaussian term at lowest order
$\Delta v_k$:
%%%%%%%%%%%%%%%%%%%%%%%%%%%%%%%%%%%%%%%%%%%%%%%%%%%%%%%
\begin{equation}
v_k(\nu) = v_k^{(G)}(\nu)+\Delta v_k(\nu,f_{\rm NL}).
\end{equation}
%%%%%%%%%%%%%%%%%%%%%%%%%%%%%%%%%%%%%%%%%%%%%%%%%%%%%%%
Each term has the following form
%%%%%%%%%%%%%%%%%%%%%%%%%%%%%%%%%%%%%%%%%%%%%%%%%%%%%%%
\begin{equation}
v_k^{(G)} = e^{-\nu^2/2}H_{k-1}(\nu),
\end{equation}
%%%%%%%%%%%%%%%%%%%%%%%%%%%%%%%%%%%%%%%%%%%%%%%%%%%%%%%
\begin{eqnarray}
\label{eq:delmf_pb}
 \Delta v_k(\nu,f_{\rm NL}) & = &
e^{-\nu^2/2}\left\{\left[
\frac16S^{(0)}H_{k+2}(\nu)+\frac{k}3S^{(1)}H_k(\nu)
\right.\right. \nonumber \\ & + &
\left.\left.\frac{k(k-1)}6S^{(2)}H_{k-2}(\nu)\right]\sigma_0 + {\cal
O}(\sigma_0^2)\right\},
\end{eqnarray}
%%%%%%%%%%%%%%%%%%%%%%%%%%%%%%%%%%%%%%%%%%%%%%%%%%%%%%%
where $H_n(\nu)$ represent the $n$-th Hermite polynomials and the
skewness parameters $S^{(k)}$ are given in Equations [27-29] of
\citet{HKM2006}. The amplitude $A_k$ (eq. [\ref{eq:mfamp}]) is not
directly relevant to non-Gaussianity but is dependent on the shape
of $C_l$. We therefore concentrate on the non-Gaussian term $\Delta
v_k$ hereafter. The quantity $\Delta v_k$ is the same as the
relative difference of MFs, which are plotted in Fig. 2 in
\citet{HKM2006}, except for its normalization factor; in this paper
the difference of MFs is normalized by $A_k$ (\ref{eq:mfamp}), while
the maximum value of MFs for Gaussian fields is used in
\citet{HKM2006}.

\subsection{Comparison with Non-Gaussian Simulations}

The above analytical formulae have already been found to match
accurately the MFs for non-Gaussian maps in Sachs-Wolfe limit
\citep[Appendix C in][]{HKM2006}. Here we test them against
non-Gaussian simulations including the full radiative transfer
function \citep{Liguori2003,Liguori2007}. As we mentioned in the
introduction, actual observations of CMB also involve different
effects which may produce other confusions: the pixel window
function, beam smearing, the inhomogeneous noise, survey mask and so
on. We include these observational effects into the simulations to
check whether they could have a systematic effect on our topological
measures.

The cosmology in the non-Gaussian simulations is based on Lambda
CDM, but the cosmological parameters have slightly different values
from WMAP three-year best-fit; $\Omega_b=0.05, \Omega_{\rm
cdm}=0.25, \Omega_\Lambda=0.7, H_0=65~{\rm km~s^{-1}~Mpc^{-1}},
\tau=0$, and $n_s=1$.  The amplitude of primordial fluctuations is,
however, set to be same as WMAP three-year best-fit value
$l(l+1)C_l/(2\pi)=5617.05({\mu}{\rm K})^2$.

Observational effects related to WMAP data are included as follows.
First we convolve the original simulation maps with the Q+V+W
co-added beam transfer function with inverse weight of the full-sky
averaged pixel-noise variance in each DA. Next we add independent
Gaussian noise realizations following the noise pattern co-added
with the same weight.  The simulation map is then masked with the
$Kp0$ Galaxy mask. Finally we smooth the simulation maps using a
Gaussian filter with a smoothing scale of $\theta_s$,
%%%%%%%%%%%%%%%%%%%%%%%%%%%%%%%%%%%%%%%%%%%%%%%%%%%%%%%
\begin{equation}
\label{eq:gaussfil}
W_l=\exp\left[-\frac12l(l+1)\theta_s^2\right].
\end{equation}
%%%%%%%%%%%%%%%%%%%%%%%%%%%%%%%%%%%%%%%%%%%%%%%%%%%%%%%
The MFs are sensitive to the resolution (or smoothing) scale of a
density field and thereby we can obtain a variety of information
from density fields using different levels of smoothing.  The
information extracted from varying smoothing scales is nevertheless
limited because they are all derived from the same original field;
the smoothed fields are not independent. Here we focus on the field
smoothed by three different smoothing scales 10', 20' and 40', where
the limit on $f_{\rm NL}$ is sufficiently converged. To remove the
effect of the survey mask near the boundary of the mask, we only use
the pixels more than $2\theta_s$ away from the boundary. The sky
fraction used in the analysis for each smoothing scale is 41\% for
$\theta_s=40'$, 62\% for $\theta_s=20'$ and 73\% for $\theta_s=10'$.

The MFs for the measured CMB temperature anisotropy are computed
from the integral of the curvature of iso-temperature contour
lengths \citep[the details are described in Appendix A.1.
of][]{HKM2006}. The binning range of $\nu$ is set to be $-3.6$ to
$3.6$ with $18$ equally spaced bins of $\nu$ per each MF. This
binning way produces well converged results irrespective of other
choices of the range of $\nu$ and the number of bins.

We obtain the normalized MFs (eq. [\ref{eq:mfamp}]) with the amplitude
$A_k$ computed from $C_l$ of each realization. Then the residuals of
the normalized MFs from Gaussian predictions $\Delta\tilde{v}_\alpha$
are calculated at each bin of $\alpha$, which denote a threshold
value $\nu$, a kind of MF $k$, and a smoothing scale
parameterized with $\theta_s$ or $N_{\rm side}$.  Even when the MFs of
Gaussian realizations are computed, however, the function
$\Delta\tilde{v}_\alpha$ are not exactly equal to $0$ due to the
effect of pixelization, survey mask and other numerical artifacts. We
therefore measure the deviations from the average of the measurements
over Gaussian realizations and subtract them as
%%%%%%%%%%%%%%%%%%%%%%%%%%%%%%%%%%%%%%%%%%%%%%%%%%%%%%%
\begin{equation}
\label{eq:mfsub}
\Delta v_\alpha=\Delta \tilde{v}_\alpha-
[\Delta \tilde{v}_\alpha]_{\rm Gaussian,mean}.
\end{equation}
%%%%%%%%%%%%%%%%%%%%%%%%%%%%%%%%%%%%%%%%%%%%%%%%%%%%%%%

\begin{figure*}
\begin{center}
\includegraphics[width=7.5cm]{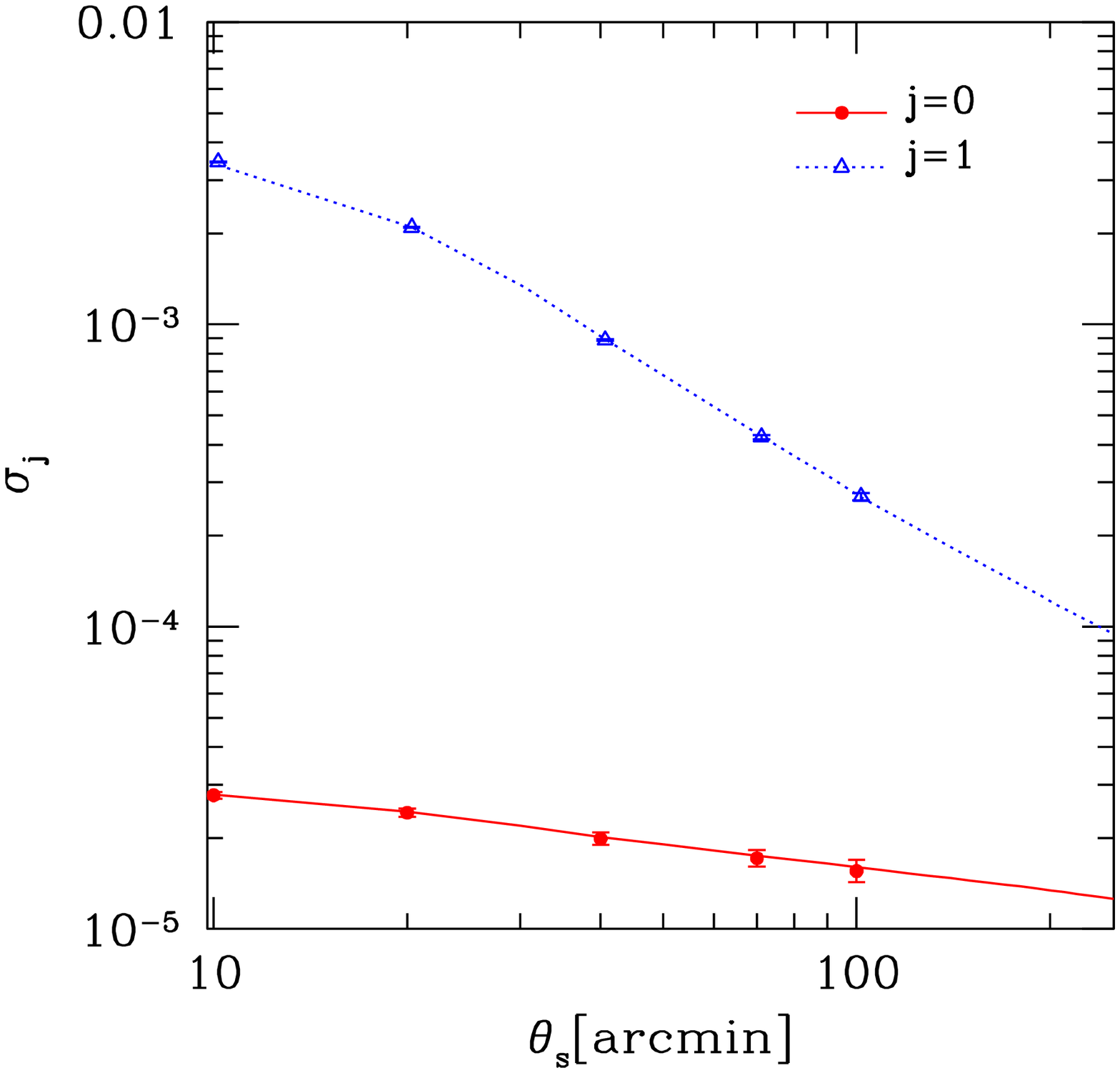}
\includegraphics[width=7.5cm]{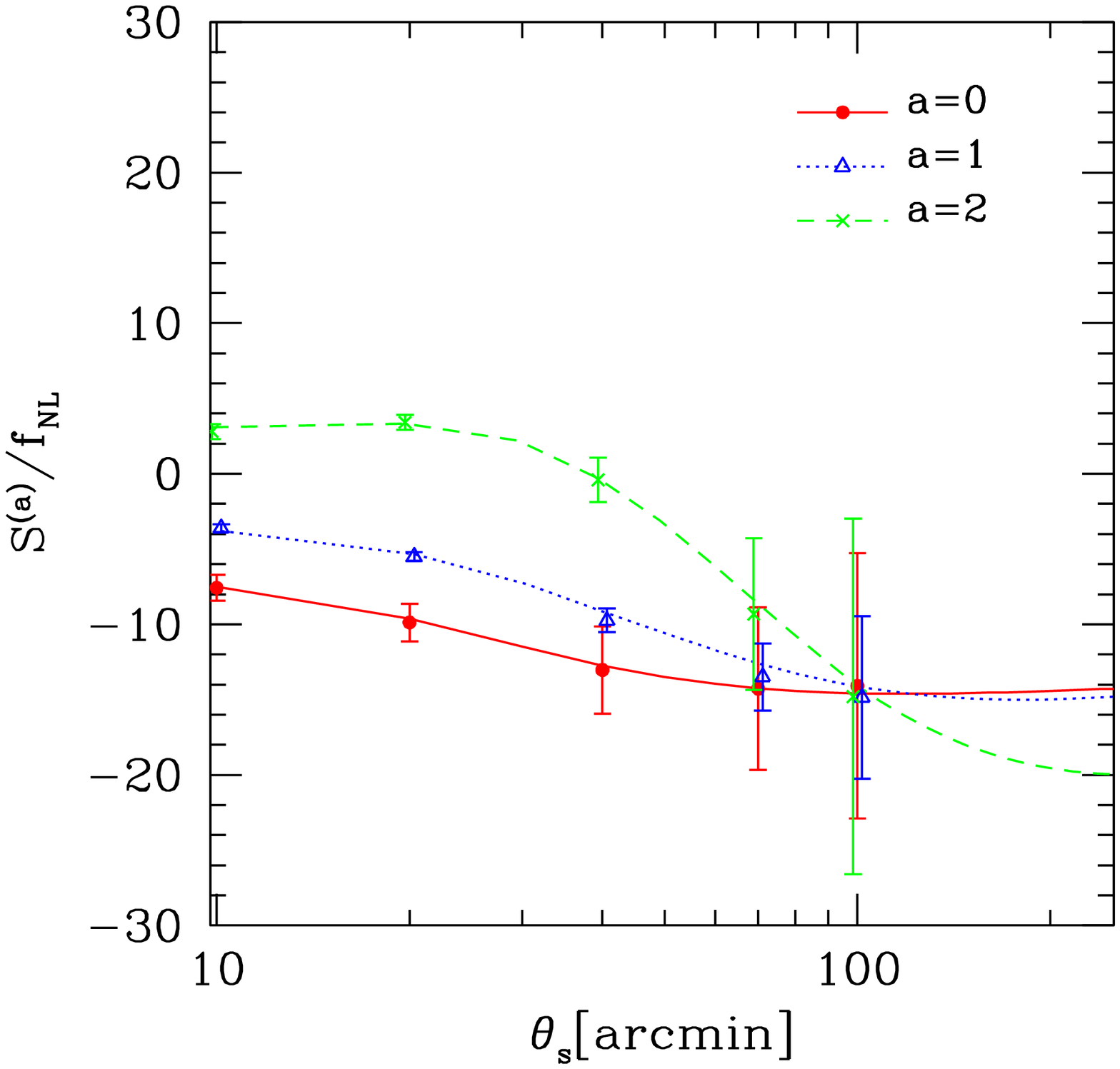}
\includegraphics[width=15cm]{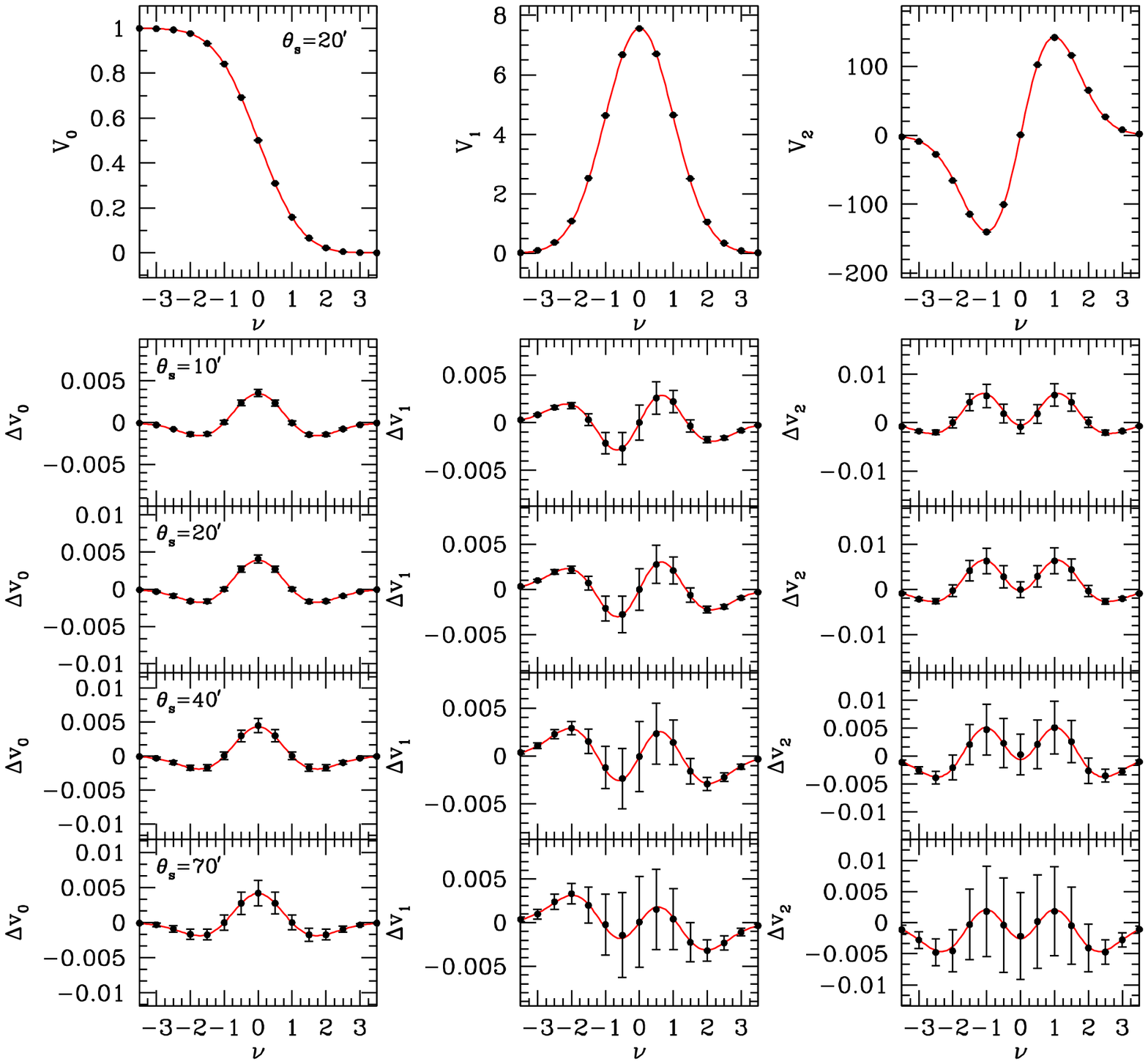}
\caption{Comparison between the analytical predictions ({\it lines})
and the numerical estimations averaged over 200 realizations
of non-Gaussian simulation maps ({\it
symbols}) for $f_{\rm NL}=100$; {\it Upper-left:} variances $\sigma_0$
and $\sigma_1$ (eq. [\ref{eq:var}]) {\it Upper-right:} skewness
parameters $S^{(a)}$ ($a=0, 1,$ and $2$) in the equation
(\ref{eq:delmf_pb}), {\it Middle:} MFs for non-Gaussian fields, $V_k$
(eq. [\ref{eq:mf}]), {\it Lower:} the difference ratio of MFs
$\Delta v_k$ (eq. [\ref{eq:delmf_pb}]). CMB maps are smoothed with
a Gaussian kernel $W_l=\exp[-l(l+1)\theta_s^2/2]$ where $\theta_s$
denotes the smoothing scale.  The fully radiative transfer function is
considered for both the theoretical predictions and the simulations. The
simulations also include the various observational effects for WMAP
three-year Q+V+W coadded map; pixel window function, beam smearing,
inhomogeneous noise pattern, and {\it Kp0} cut. The error-bars represent
the errors for the averaged simulation results over 200 realizations
(the sample variance divided by the square-root-of 200).}
\label{fig:cmbsim}
\end{center}
\end{figure*}

In Fig. \ref{fig:cmbsim}, we compare the analytical predictions of
variance, skewness, and MFs with the measurements from the simulations
for $f_{\rm NL}=100$. The simulated results are the average over 200
realizations and the error-bars represent the error for the average
(the sample variance divided by the square-root-of $200$).  
The averaged measurements for Gaussian CMB maps are subtracted from those
for non-Gaussian maps in the simulated plots including variance and
skewness as well as MFs (see eq. [\ref{eq:mfsub}]). In the
plots, we adopt the Gaussian maps which are generated from the same
realizations of linear potential fields ($\phi$ in eq. [1]) as
the non-Gaussian CMB maps.  The sample variances, represented by the
error-bars, are cancelled very well in such plots so one can focus on
the systematic effect of primordial enon-Gaussianity.  The analytical
formulae are found to agree with the simulations extremely well even
including all observational effects. This indicates that we can
measure $f_{\rm NL}$ from direct comparison of the analytical formulae
with observations without having to worry excessively about the
presence of such systematics. We also check that both the artificial
systematics $[\Delta \tilde{v}_\alpha]_{\rm Gaussian,mean}$ and
covariance matrix are not strongly dependent on the details of
cosmology. These results are encouraging, but not unexpected: being
based on integrated properties, the Minkowski Functionals are expected
to be robust to such effects.

%%%%%%%%%%%%%%%%%%%%%%%%%%%%%%%%%%%%%%%%%%%%%%%%%%%%%%%
%%%%%%%%%%%%%%%%%%%%%%%%%%%%%%%%%%%%%%%%%%%%%%%%%%%%%%%

\section{Constraints on Primordial Non-Gaussianity
from WMAP Three-Year Data}
\label{sec:wmaplimit}

\subsection{Covariance Matrix for MFs}
%%%%%%%%%%%%%%%%%%%%%%%%%%%%%%%%%%%%%%%%%%%%%%%%%%%%%%%
\begin{figure*}
\begin{center}
\includegraphics[width=8cm]{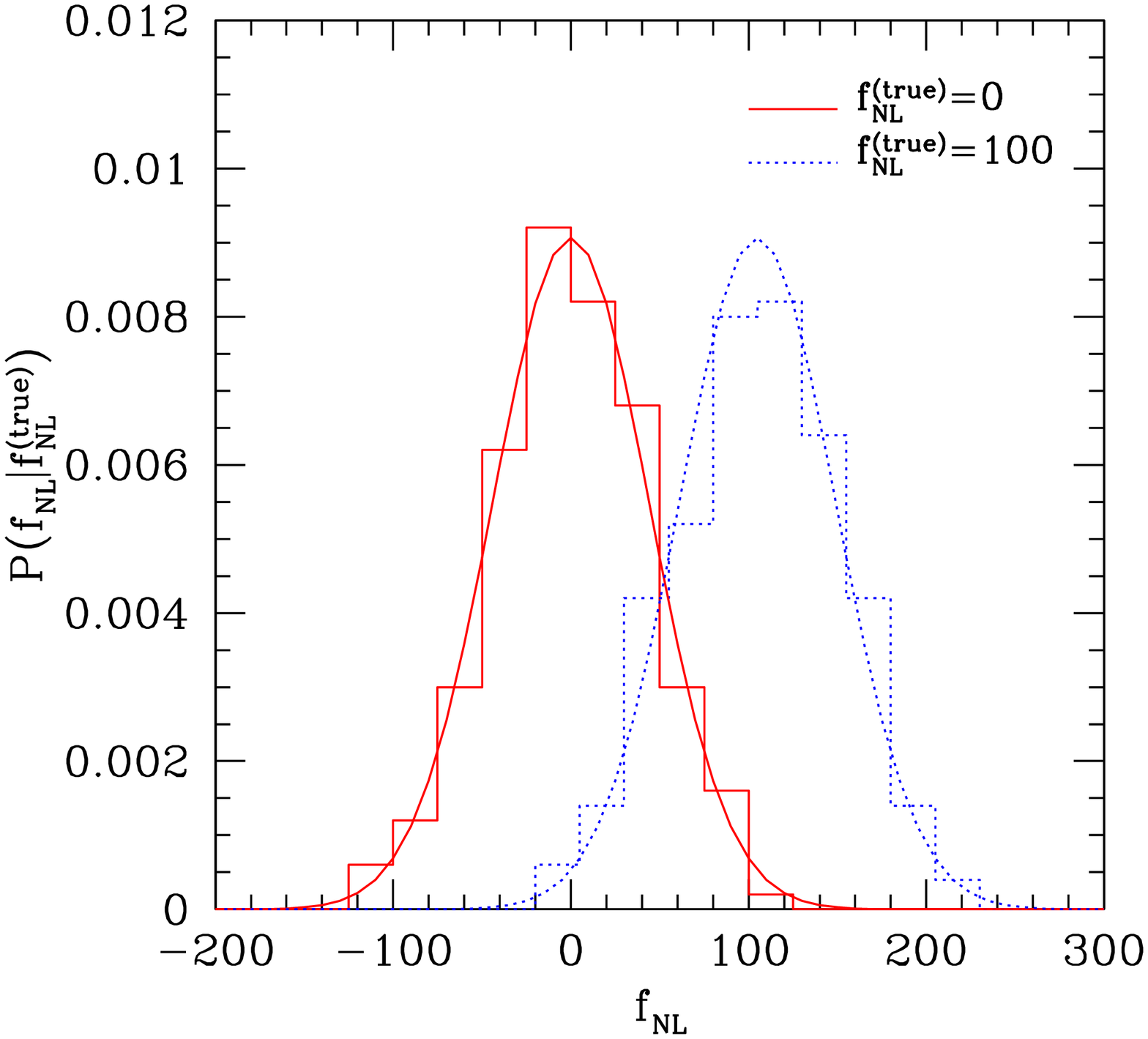}
\caption{The distribution function of the best-fit value of $f_{\rm
NL}$ using WMAP three-year mock simulation maps (histogram).  We use
$200$ realizations of Gaussian simulations (solid) and non-Gaussian
simulations with $f_{\rm NL}^{\rm (true)}=100$ (dotted)
respectively. The best-fit values are obtained by fitting the
analytical formulae (eq. [\ref{eq:delmf_pb}]) to all of the MFs for
the simulations at $\theta_s=10'$ and $20'$ combined.  For comparison,
we plot the likelihood function of $f_{\rm NL}$ at $f_{\rm NL}^{\rm
(true)}=0$ and 100 with $\sigma_{f_{\rm NL}}=44$
(eq.  [\ref{eq:likefunc_fnl}]), which is the expected
uncertainty of $f_{\rm NL}$ from all of the MFs for CMB maps
at $\theta_s=10'$ and $20'$ combined.
}%
\label{fig:fnlfit}
\end{center}
\end{figure*}
%%%%%%%%%%%%%%%%%%%%%%%%%%%%%%%%%%%%%%%%%%%%%%%%%%%%%%%

We have adopted a maximum likelihood method to estimate the best-fit
value of $f_{\rm NL}$ and its associated uncertainty.  In `nearly
Gaussian' fields, the distribution functions of $\Delta v_\alpha$
are well described as multivariate Gaussians. The likelihood
function of $f_{\rm NL}$ is, therefore, simply proportional to
$\exp(-\chi^2(f_{\rm NL})/2)$ where $\chi^2(f_{\rm NL})$ is computed
using the theoretical formulae (eq. [\ref{eq:delmf_pb}]) as
%%%%%%%%%%%%%%%%%%%%%%%%%%%%%%%%%%%%%%%%%%%%%%%%%%%%%%%
\begin{eqnarray}
\label{eq:covmatrix}
  \chi^2(f_{\rm NL})&=&\sum_{\alpha\alpha'}
  [\Delta v_\alpha^{\rm (obs)}-\Delta v_\alpha^{\rm (theory)}(f_{\rm NL})]
  \Sigma^{-1}_{\alpha\alpha'} \nonumber \\
  & \times &
  [\Delta v_{\alpha'}^{\rm (obs)}-\Delta v_{\alpha'}^{\rm
(theory)}(f_{\rm NL})],
\end{eqnarray}
%%%%%%%%%%%%%%%%%%%%%%%%%%%%%%%%%%%%%%%%%%%%%%%%%%%%%%%
where $\alpha$ and $\alpha^\prime$ denote the binning number of
threshold values $\nu$, different kinds of MF $k$, and smoothing scale
parameterized with $\theta_s$ or $N_{\rm side}$.  The full covariance
matrix $\Sigma_{\alpha\alpha'}$ is required because MFs are strongly
correlated between different $\nu$, different kinds of MF, and also
different $N_{\rm side}$ or $\theta_s$. We estimate the covariance
matrix of MFs from 1000 Gaussian simulation maps including the pixel
and beam window function, {\it Kp0} survey mask, and inhomogeneous
noise for WMAP three-year maps.

The MFs contain information about fluctuations at different scales,
so the results depend on the choice of window function. Here, two
different types of window functions are adopted (in addition to the
beam window functions). One is a Gaussian window function with the
scale characterized by $\theta_s$ (which is chosen to be
sufficiently large compared with the pixel size). The other is just
the pixel window function in HEALPix format with a scale
characterized by $N_{\rm side}$.  The multipole components with $l$
higher than $2N_{\rm side}$ are cut because they suffer from serious
aliasing effects.

Before applying these ideas to the observational data, we check if
our method based on $\chi^2$ analysis is valid using simulations
with primordial non-Gaussianity. We apply the non-Gaussian
simulations to check that the likelihood function using the equation
(\ref{eq:covmatrix}) reproduces a valid probability distribution of
the true value $f_{\rm NL}^{\rm (true)}$. Here we consider that the
likelihood function of $f_{NL}^{\rm (true)}$ in each realization
follows a Gaussian distribution around the best-fit value
$f_{NL}^{\rm (best)}$ as
%%%%%%%%%%%%%%%%%%%%%%%%%%%%%%%%%%%%%%%%%%%%%%%%%%%%%%%
\begin{eqnarray}
\label{eq:likefunc_fnl}
P(f_{\rm NL}^{\rm (true)}|f_{\rm NL}^{\rm (best)})
&=&\frac{1}{\sqrt{2\pi}\sigma_{f_{\rm NL}}}
\exp\left[-\frac{1}{2}
\left(\frac{f_{\rm NL}^{\rm (true)}-f_{\rm NL}^{\rm
(best)}}{\sigma_{f_{\rm NL}}}\right)^2\right] \nonumber \\
& & \\
\sigma_{f_{\rm NL}}&=&\left[\sum_{\alpha\alpha'}\frac{\partial
  V_\alpha}{\partial
  f_{\rm NL}}(\Sigma^{-1})_{\alpha\alpha'}\frac{\partial
  V_{\alpha'}}{\partial f_{\rm NL}}\right]^{-1/2}
\end{eqnarray}
%%%%%%%%%%%%%%%%%%%%%%%%%%%%%%%%%%%%%%%%%%%%%%%%%%%%%%%
where the binning number $\alpha$ represents a threshold $\nu$ for
$k$-th MF at a given scale $N_{\rm side}$ (or $\theta_s$ if the
Gaussian smoothing is added) and the covariance matrix
$\Sigma_{\alpha\alpha'}$ is numerically estimated from Gaussian
simulations.  The function $\partial V_\alpha/\partial f_{\rm NL}$
is independent of $f_{\rm NL}$ (see equation [\ref{eq:delmf_pb}])
and thus the uncertainty $\sigma_{f_{\rm NL}}$ is independent of
$f_{\rm NL}$.  According to Bayes' theorem, $f_{\rm NL}^{\rm
(best)}$ should distribute around $f_{\rm NL}^{\rm (true)}$ in the
same way:
%%%%%%%%%%%%%%%%%%%%%%%%%%%%%%%%%%%%%%%%%%%%%%%%%%%%%%%
\begin{equation}
P(f_{\rm NL}^{\rm (best)}|f_{\rm NL}^{\rm (true)})=
P(f_{\rm NL}^{\rm (true)}|f_{\rm NL}^{\rm (best)})
\end{equation}
%%%%%%%%%%%%%%%%%%%%%%%%%%%%%%%%%%%%%%%%%%%%%%%%%%%%%%%

We estimate the distribution function of $f_{\rm NL}^{\rm (best)}$
from 200 non-Gaussian CMB simulated maps at a given $f_{\rm NL}^{\rm
(true)}$ and then compare with the equation (\ref{eq:likefunc_fnl}).
The simulated maps include observational effects represented by
pixel and beam window functions, noise, and {\it Kp0} survey cut for
WMAP three-year data.  Fig. \ref{fig:fnlfit} shows the theoretical
predictions of $P(f_{\rm NL}^{\rm (best)}| f_{\rm NL}^{\rm (true)})$
at $f_{\rm NL}^{\rm (true)}=0$ (solid) and 100 (dotted).  from the
MFs for the combined maps at Gaussian smoothing scales
$\theta_s=20'$ and $10'$ where the uncertainty is $\sigma_{f_{\rm
NL}}=44$.  The histograms show the distribution of $f_{\rm NL}^{\rm
(best)}$ from 200 (non-)Gaussian realizations. The averages of the
best fit values of $f_{\rm NL}$ from the simulations are
respectively $0\pm 44$ (for $f_{\rm NL}^{\rm (true)}=0$) and $101\pm
46$ (for $f_{\rm NL}^{\rm (true)}=100$).  The simulations reproduce
the theoretical predictions of the likelihood function very well.
Our method is thus well established to give constraints on $f_{\rm
NL}$ from WMAP three-year map.

\subsection{Constraints on $f_{\rm NL}$ from Minkowski Functionals for WMAP Three-Year Temperature Anisotropy}

\begin{figure*}
\begin{center}
\includegraphics[width=8.5cm]{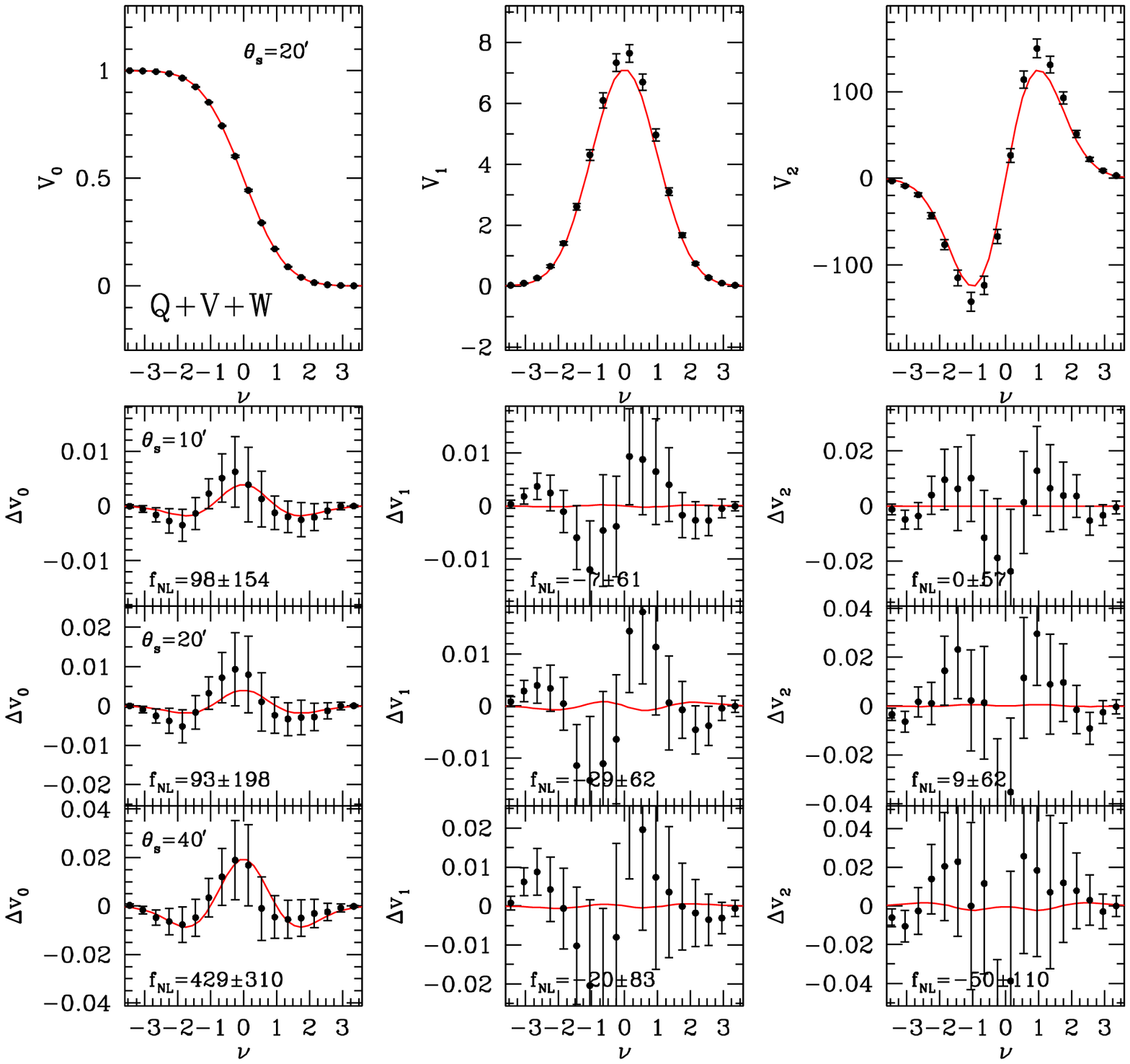}
\includegraphics[width=8.5cm]{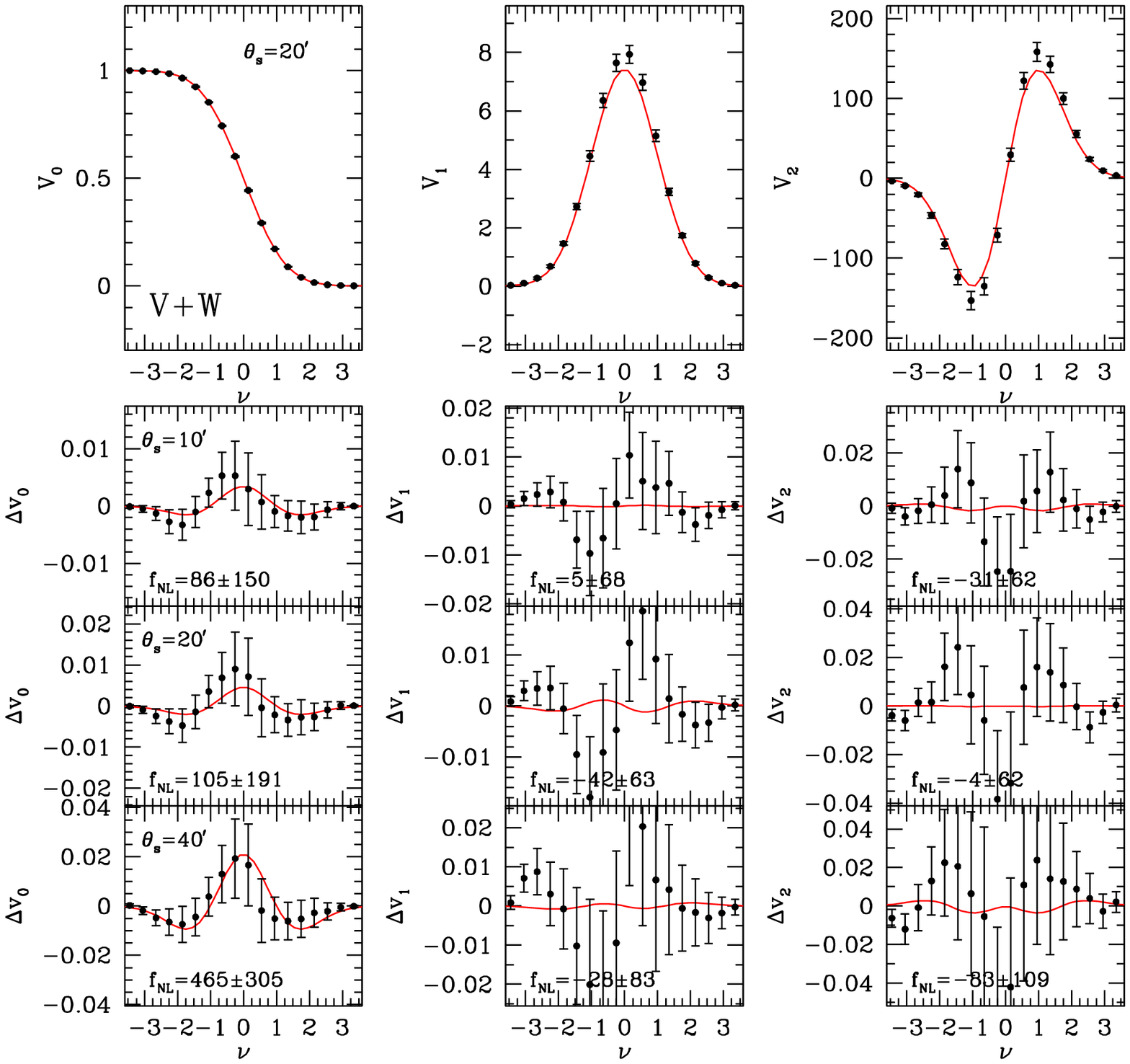}
\caption{Comparison between MFs for WMAP three-year temperature maps
(symbols) and the analytical formulae with the best-fit value of
$f_{\rm NL}$ for each MF (lines).  The MFs
are calculated from the Q+V+W co-added map ({\it Left}) and the V+W
map ({\it Right}). Top panels show the MFs $V_k$ ($k$=0,1, and 2) at
$\theta_s$=20 arcmin, and other panels illustrate $\Delta v_k$
(eq. [\ref{eq:delmf_pb}]) at $\theta_s$=10, 20 and 40 arcmin
respectively. Error-bars denote the standard deviation of MFs at each
bin of $\nu$ computed from 1000 Gaussian realizations including the
WMAP three-year noise distribution, {\it Kp0} mask and pixel and beam
window function. The systematics due to the pixelization effect is
estimated from the Gaussian realizations and is subtracted from the
observed MFs (see eq. [\ref{eq:mfsub}]).}
\label{fig:mf3year_GA}
\end{center}
\end{figure*}

\begin{figure*}
\begin{center}
\includegraphics[width=8.5cm]{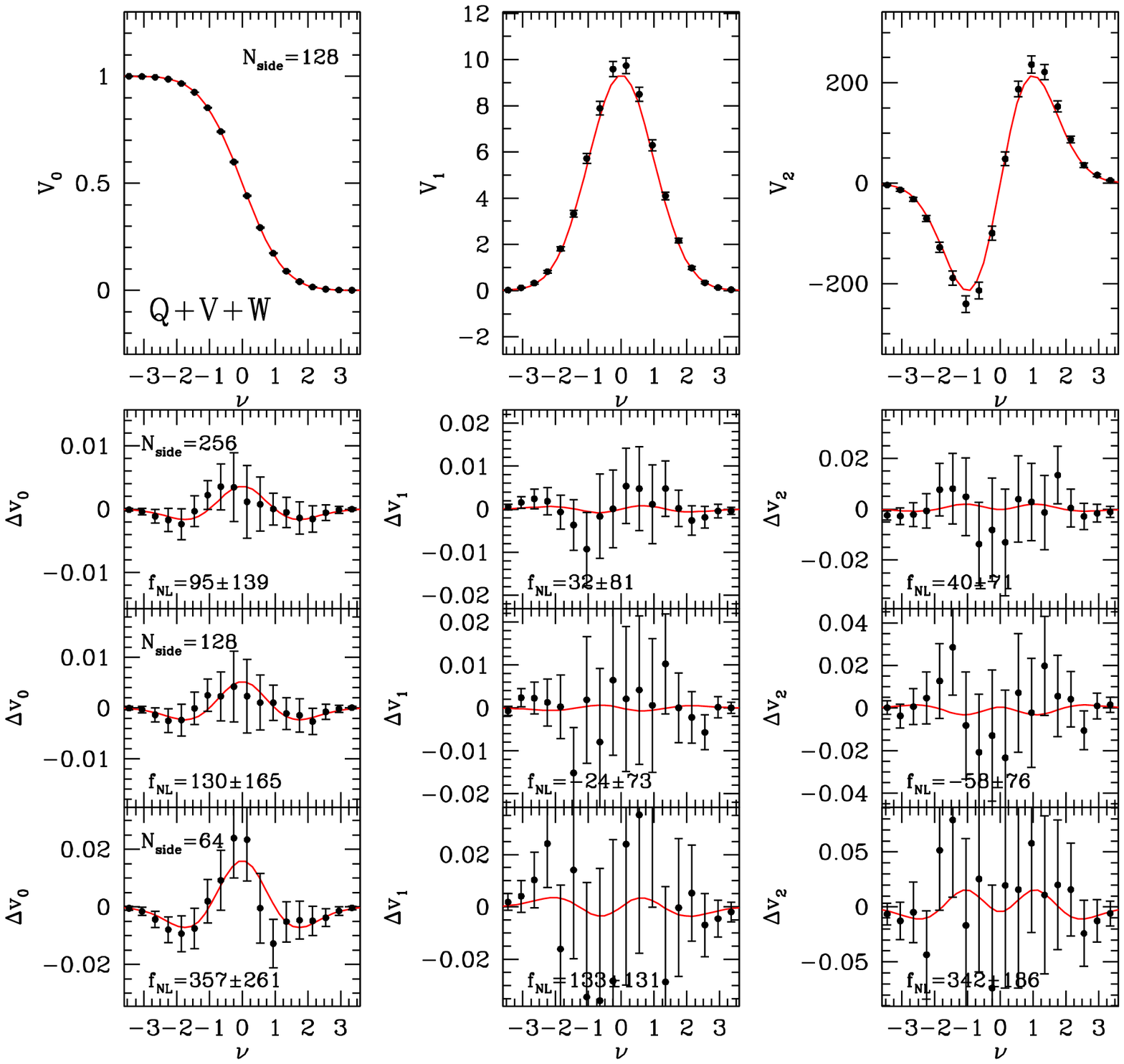}
\includegraphics[width=8.5cm]{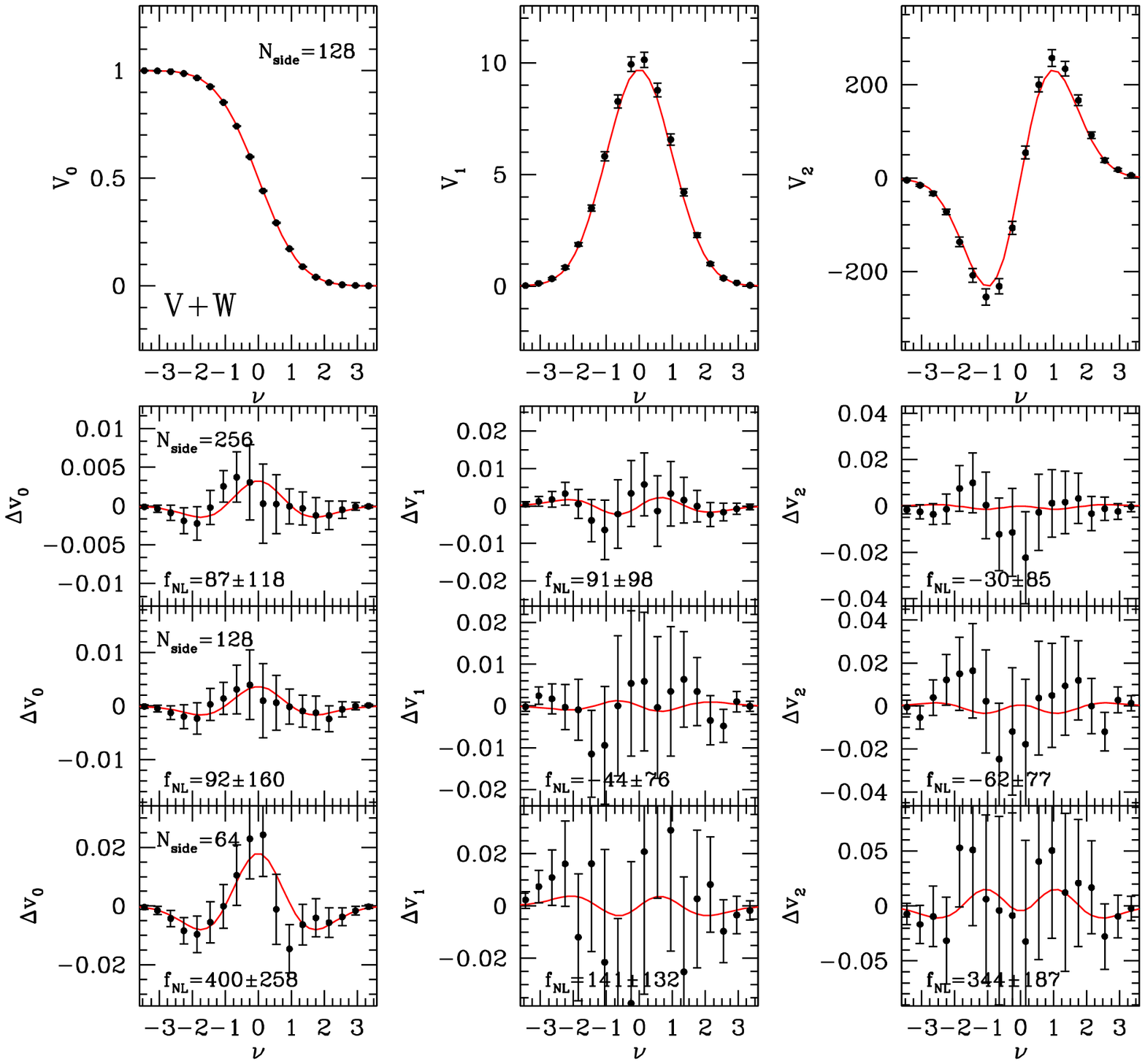}
\caption{Same as Fig. \ref{fig:mf3year_GA} but for different $N_{\rm
side}=256, 128$ and $64$ without Gaussian smoothing.}
\label{fig:mf3year_TH}
\end{center}
\end{figure*}

The three MFs for the CMB temperature maps from WMAP three-year data
are respectively plotted with symbols in each column of
Fig. \ref{fig:mf3year_GA} (left three columns for the ``Q+V+W'' map
and right three columns for the ``V+W'' map).  In each column, the top
panel shows the MF $V_k$ at a representative scale ($\theta_s=20'$).
and the lower three panels illustrate $\Delta v_k$ at different
$\theta_s=10',20'$ and $40'$.  The perturbative formulae with the
best-fit value of $f_{\rm NL}$ to each observed MF are plotted with
lines. The best-fit values and $1\sigma$ uncertainty are written in
the left-bottom side of each panel.  In top panels, all of the
amplitude of observed MFs are found to be smaller than the theoretical
estimations. This comes from the deficit of the observed power at low
$l$ which generates the larger amplitude of MFs determined by
$\sigma_1/\sigma_0$ (eq.[\ref{eq:mfamp}]), as pointed out by
\citet{Gott2007}.  Fig. \ref{fig:mf3year_TH} shows the same plot but
for the MFs with the pixel window function only.  The top panel shows
the MF at $N_{\rm side}=128$ and the lower three panels illustrate
$\Delta v_k$ for $N_{\rm side}$=256, 128, and 64.  It is interesting
that all MFs at $N_{\rm side}$=64 have large positive values of
$f_{\rm NL}$, though the significance is less than $2\sigma$.

Table \ref{tab:fnl3year_GA} lists the best-fit values and the
$1\sigma$ uncertainty of $f_{\rm NL}$ for each MF and their combined
values at different sets of Gaussian smoothing scales $\theta_s$.
The $1\sigma$ uncertainty of $f_{\rm NL}$ is estimated from the
range of $f_{\rm NL}$ with $\Delta\chi^2=\chi^2-\chi^2_{\rm min}\le
1$.  The minimum of chi-square $\chi_{\rm min}$ and the
goodness-of-fit $P_{\chi^2>\chi^2_{\rm min}}$ are listed for each
fit.  The results for the pixel window function only are shown in
Table \ref{tab:fnl3year_TH}. The goodness-of-fit values are
reasonable  for all the fits, which means that the simple form of
the primordial non-Gaussianity (equation [\ref{eq:ngpotential2}])
well describes the behaviour of the observed MFs. In other words,
present observations are too uncertain to allow the extraction of
any further information about primordial non-Gaussianity (e.g. scale
dependence of $f_{\rm NL}$). The constraint $-70<f_{\rm NL}<91$ at
95\% C.L. is obtained from all MFs for the Q+V+W co-added map at
combined different Gaussian smoothing scales of 10, 20 and 40
arcmin. A similar constraint is obtained from the MFs with
pixel-window only as $-84<f_{\rm NL}<105$. The results from Q+V+W
co-added map are consistent with the previous ones
\citep{Spergel2007,Creminelli2007a}.

There is some friction (but not disagreement) between our results and
those by \citet{YW2008}; our V+W analysis finds $f_{\rm NL}=-22\pm 43$
whereas they find $f_{\rm NL}=87\pm 30$.  Moreover our averaged
$f_{\rm NL}$ decreases from Q+V+W to V+W whereas their $f_{\rm NL}$
increases. This is very interesting because there is the possibility
that e.g., foregrounds and point sources might be biasing one of the
two results. \citet{YW2008} show in their analysis that these effects
do not seem to contaminate the primordial bispectrum measurement
significantly. It will be then important to check their effect on the
MFs statistics in order to verify if this can explain the differencies
among the two results.  However the observed discrepancies show
already how analyzing non-Gaussianity using different statistics can
provide additional interesting information.

%%%%%%%%%%%%%%%%%%%%%%%%%%%%%%%%%%%%%%%%%%%%%%%%%%%%%%%
%%%%%%%%%%%%%%%%%%%%%%%%%%%%%%%%%%%%%%%%%%%%%%%%%%%%%%%
%%%%%%%%%%%%%%%%%%%%%%%%%%%%%%%%%%%%%%%%%%%%%%%%%%%%%%%

%%%%%%%%%%%%%%%%%%%%%%%%%%%%%%%%%%%%%%%%%%%%%%%%%%%%%%%
\begin{table*}
\caption{The constraints on $f_{\rm NL}$ from MFs for WMAP three-year
co-added maps, Q+V+W and V+W, at different Gaussian smoothing scales
$\theta_s$ [arcmin] of 40, 20 and 10 and their combination. For the
calculation of their constraints, the maximum likelihood method are
employed with the analytical formulae (eq. [\ref{eq:delmf_pb}]) and the
covariance matrix of MFs estimated from 1000 Gaussian realizations.
The range of $\nu$ is set from $-3.6$ to $3.6$ with the binning number per
each MF of 18. The goodness-of-fit of the analytical formulae is represented by
the minimum values of $\chi^2$ value, $\chi^2_{\rm min}$, and the probability
with $\chi^2$ larger than $\chi^2_{\rm min}$.}
\begin{center}
\begin{tabular}{ccccccc}
  \hline\hline
 & & & \multicolumn{2}{c}{Q+V+W} & \multicolumn{2}{c}{V+W} \\
\cline{4-7}
\raisebox{1.5ex}[0pt]{$\theta_s$ [arcmin]} &
\raisebox{1.5ex}[0pt]{MF} &
\raisebox{1.5ex}[0pt]{d.o.f.} &
\raisebox{-1ex}{$\chi^2_{\rm min}(P_{\chi^2>\chi^2_{\rm min}})$} &
\raisebox{-1ex}{$f_{\rm NL}$} &
\raisebox{-1ex}{$\chi^2_{\rm min}(P_{\chi^2>\chi^2_{\rm min}})$} &
\raisebox{-1ex}{$f_{\rm NL}$} \\ \hline
            40 & $V_0$  &    17 &  19.2 (0.32) & $ 429\pm 310$ &  23.7 (0.13) & $ 465\pm 305$\\
            40 & $V_1$  &    17 &  12.9 (0.75) & $ -20\pm  82$ &  16.7 (0.48) & $ -28\pm  83$\\
            40 & $V_2$  &    17 &   8.2 (0.96) & $ -50\pm 109$ &   7.6 (0.97) & $ -83\pm 109$\\
            40 & All    &    53 &  46.8 (0.71) & $   7\pm  76$ &  48.8 (0.64) & $ -14\pm  77$\\ \hline
            20 & $V_0$  &    17 &  20.8 (0.24) & $  93\pm 198$ &  20.7 (0.24) & $ 105\pm 191$\\
            20 & $V_1$  &    17 &  14.0 (0.66) & $ -29\pm  62$ &  12.6 (0.76) & $ -42\pm  63$\\
            20 & $V_2$  &    17 &  13.9 (0.67) & $   9\pm  62$ &  10.9 (0.86) & $  -4\pm  62$\\
            20 & All    &    53 &  47.9 (0.67) & $ -32\pm  48$ &  49.4 (0.61) & $ -53\pm  49$\\ \hline
            10 & $V_0$  &    17 &   9.0 (0.94) & $  98\pm 154$ &   9.4 (0.93) & $  86\pm 149$\\
            10 & $V_1$  &    17 &  11.8 (0.81) & $  -7\pm  61$ &  11.3 (0.84) & $   5\pm  67$\\
            10 & $V_2$  &    17 &   9.9 (0.91) & $   0\pm  57$ &   7.9 (0.97) & $ -31\pm  61$\\
            10 & All    &    53 &  42.4 (0.85) & $ -10\pm  46$ &  52.7 (0.48) & $ -25\pm  52$\\ \hline
  10,~20~\&~40 & $V_0$  &    53 &  49.0 (0.63) & $   8\pm  74$ &  55.5 (0.38) & $ -20\pm  76$\\
  10,~20~\&~40 & $V_1$  &    53 &  41.4 (0.88) & $ -20\pm  55$ &  44.2 (0.80) & $ -23\pm  57$\\
  10,~20~\&~40 & $V_2$  &    53 &  34.8 (0.98) & $   5\pm  52$ &  28.3 (1.00) & $ -19\pm  53$\\
  10,~20~\&~40 & All    &   161 & 148.4 (0.75) & $  11\pm  40$ & 173.2 (0.24) & $ -22\pm  43$\\ \hline
\end{tabular}
\end{center}
\label{tab:fnl3year_GA}
\end{table*}
%%%%%%%%%%%%%%%%%%%%%%%%%%%%%%%%%%%%%%%%%%%%%%%%%%%%%%%
%%%%%%%%%%%%%%%%%%%%%%%%%%%%%%%%%%%%%%%%%%%%%%%%%%%%%%%
\begin{table*}
\caption{Same as Table \ref{tab:fnl3year_GA} but for
different $N_{\rm side}=64, 128$ and $256$ without Gaussian smoothing}
\begin{center}
\begin{tabular}{ccccccc}
  \hline\hline
 & & & \multicolumn{2}{c}{Q+V+W} & \multicolumn{2}{c}{V+W} \\
\cline{4-7}
\raisebox{1.5ex}[0pt]{$N_{\rm side}$} &
\raisebox{1.5ex}[0pt]{MF} &
\raisebox{1.5ex}[0pt]{d.o.f.} &
\raisebox{-1ex}{$\chi^2_{\rm min}(P_{\chi^2>\chi^2_{\rm min}})$} &
\raisebox{-1ex}{$f_{\rm NL}$} &
\raisebox{-1ex}{$\chi^2_{\rm min}(P_{\chi^2>\chi^2_{\rm min}})$} &
\raisebox{-1ex}{$f_{\rm NL}$} \\ \hline
            64 & $V_0$  &    17 &  17.5 (0.42) & $ 357\pm 260$ &  18.7 (0.34) & $ 400\pm 257$\\
            64 & $V_1$  &    17 &  18.7 (0.34) & $ 133\pm 131$ &  19.2 (0.32) & $ 141\pm 132$\\
            64 & $V_2$  &    17 &  12.2 (0.79) & $ 342\pm 186$ &   8.3 (0.96) & $ 344\pm 187$\\
            64 & All    &    53 &  46.9 (0.71) & $ 137\pm 122$ &  44.1 (0.80) & $ 104\pm 123$\\ \hline
           128 & $V_0$  &    17 &  25.0 (0.09) & $ 130\pm 165$ &  15.5 (0.56) & $  92\pm 160$\\
           128 & $V_1$  &    17 &  25.5 (0.08) & $ -24\pm  73$ &  19.7 (0.29) & $ -44\pm  76$\\
           128 & $V_2$  &    17 &  14.1 (0.66) & $ -58\pm  76$ &  16.5 (0.49) & $ -62\pm  77$\\
           128 & All    &    53 &  55.8 (0.37) & $ -10\pm  60$ &  41.6 (0.87) & $ -45\pm  62$\\ \hline
           256 & $V_0$  &    17 &   7.2 (0.98) & $  95\pm 139$ &  13.1 (0.73) & $  87\pm 118$\\
           256 & $V_1$  &    17 &   8.7 (0.95) & $  32\pm  81$ &   9.4 (0.93) & $  91\pm  98$\\
           256 & $V_2$  &    17 &  10.1 (0.90) & $  40\pm  71$ &   9.4 (0.93) & $ -30\pm  85$\\
           256 & All    &    53 &  38.1 (0.94) & $  26\pm  60$ &  41.0 (0.89) & $ -13\pm  69$\\ \hline
256,~128~\&~64 & $V_0$  &    53 &  55.9 (0.37) & $ -44\pm 109$ &  54.5 (0.42) & $ -21\pm  99$\\
256,~128~\&~64 & $V_1$  &    53 &  61.4 (0.20) & $ -15\pm  63$ &  55.2 (0.39) & $ -20\pm  67$\\
256,~128~\&~64 & $V_2$  &    53 &  41.5 (0.87) & $  -7\pm  60$ &  41.2 (0.88) & $ -51\pm  63$\\
256,~128~\&~64 & All    &   161 & 165.3 (0.39) & $  11\pm  47$ & 152.4 (0.67) & $ -48\pm  48$\\ \hline
\end{tabular}
\end{center}
\label{tab:fnl3year_TH}
\end{table*}
%%%%%%%%%%%%%%%%%%%%%%%%%%%%%%%%%%%%%%%%%%%%%%%%%%%%%%%
%\clearpage

\section{Summary and Conclusions}
\label{sec:summary}

We have presented an analysis of MFs for WMAP the three-year
temperature maps to limit the primordial non-Gaussianity
characterized by the nonlinear coupling parameter $f_{\rm NL}$. To
do this we compared perturbative formulae for MFs of weakly
non-Gaussian fields directly with the observations.  The analytical
formulae are found to be in excellent agreement with results from
non-Gaussian simulations of CMB  maps including full radiative
transfer effects. The agreement is still very good when including
systematic observational effects including the {\it Kp0} survey
mask, pixel and beam window functions, and inhomogeneous noise
distribution for WMAP three-year data.

We have performed a $\chi^2$ analysis to the comparison of the
analytical formulae with WMAP three-year data. The fits of the
analytical formulae to the observations are acceptable and we thus
obtain a robust constraint of  $-70<f_{\rm NL}<91$ at 95\% C.L.
from the Q+V+W coadded maps with Gaussian filter at different
scales 10', 20' and 40' combined. The result is consistent
with previous results \citep{Spergel2007,Creminelli2007a,YW2008}.

The behaviour of the results for the V+W maps raises some
interesting issues; our constraint is negatively shifted
$-108<f_{\rm NL}<64$ while \citet{YW2008} find a more positive range
$27<f_{\rm NL}<147$.  The difference between the two results should
be clearer in the near future survey represented by Planck. It is
worth investigating this result in further detail through a careful
analysis of foregrounds and point source effects. This will be the
subject of  future work.

%While the consistency of this result with other estimates of, and
%constrains on, $f_{\rm NL}$ is very encouraging, there remains the
%possibility that some form of non-Gaussianity may exist but which
%has eluded the methods deployed to detect it. The quest for new
%diagnostics must continue, but in the meantime the picture that is
%emerging is satisfyingly coherent.

%%%%%%%%%%%%%%%%%%%%%%%%%%%%%%%%%%%%%%%%%%%%%%%%%%%%%%%%%%%%%%%%%%%%%%%
\section*{Acknowledgments}
We thank the anonymous referee for providing very useful comments and
suggestions. We deeply appreciate Eiichiro Komatsu who originally
proceeded with the project together.  C.~H. acknowledges support from
the Particle Physics and Astronomy Research Council grant number
PP/C501692/1. C.H. also acknowledges support from a JSPS (Japan
Society for the Promotion of Science) fellowship. T.~M. acknowledges
the support from the Ministry of Education, Culture, Sports, Science,
and Technology, Grant-in-Aid for Scientific Research (C), 18540260,
2006, and Grant-in-Aid for Scientific Research on Priority Areas
No. 467 ``Probing the Dark Energy through an Extremely Wide \& Deep
Survey with Subaru Telescope''. S.~M. acknowledges ASI contract Planck
LFI Activity of Phase E2, for partial financial support.

%%%%%%%%%%%%%%%%%%%%%%%%%%%%%%%%%%%%%%%%%%%%%%%%%%%%%%%%%%%%%%%%%%%%%%%

%%%%%%%%%%%%%%%%%%%%%%%%%%%%%%%%%%%%%%%%%%%%%%%%%%%%%%%

\end{document}